\documentclass[twocolumn,showpacs,preprintnumbers,amsmath,amssymb,superscriptaddress]{revtex4}

\newcommand{\beq}{\begin{equation}}
\newcommand{\eeq}{\end{equation}}
\def\beqa{\begin{eqnarray}}
\def\eeqa{\end{eqnarray}}
\def\hmpc{h^{-1}\,{\rm Mpc}}
\def\hkpc{h^{-1}\,{\rm kpc}}
\def\lcdm{\Lambda{\rm CDM}}

\usepackage{graphicx}
\usepackage{dcolumn}

\usepackage{bm}

\begin{document}

\title{Dark Matter Gravitational Clustering With a Long-Range Scalar Interaction}

\author{Wojciech A. Hellwing}
\email{pchela@camk.edu.pl}
\affiliation{Nicolaus Copernicus Astronomical Center, Bartycka 18, 00-719 Warsaw, Poland}
\author{Roman Juszkiewicz}
\email{roman@camk.edu.pl}
\affiliation{Nicolaus Copernicus Astronomical Center, Bartycka 18, 00-719 Warsaw, Poland}
\affiliation{Institute of Astronomy, University of Zielona G\'ora, ul. Lubuska 2, Zielona G\'ora, Poland}

\date{\today}

\begin{abstract}
We explore the possibility improving the $\Lambda$CDM model at megaparsec scales by introducing
a scalar interaction that increases the mutual gravitational attraction of dark matter particles.
Using N-body simulations, we study the spatial distribution of dark matter particles and halos.
We measure the effect of modifications in the Newton's gravity on properties of the two-point
correlation function, the dark matter power spectrum, the cumulative halo mass function and density probability distribution functions.
The results look promising: the scalar interactions produce desirable features at megaparsec scales without spoiling the $\Lambda$CDM successes
at larger scales.  

\end{abstract}
\maketitle

\section{\label{introduction}Introduction}

In this paper we study cosmological implications of a scalar interaction
that produces a long-range fifth force 
in the dark sector, proposed by G.~Farrar , S.~Gubser and J.~Peebles \cite{FarrarPeebles,GP1,GP2,Farrar2007,Brookfield}.  
The physical motivation for this model comes from the string theory \cite{StringGas}.

The cosmological motivation comes from small-scale difficulties of the $\Lambda$CDM model, 
which successfully passes almost all observational tests (see e.g. \cite{WMAP07}). 
Difficulties appear at length scales below few megaparsecs: (1) the $\Lambda$CDM voids are less empty 
than the real voids; (2) the present accretion rate of intergalactic debris onto thin spiral galaxies poses a problem for current galaxy formation paradigm;
(3) merger rates at low redshifts for giant elliptical galaxies suggest violent accretion histories for their haloes at low redshifts which is in
contradiction with observations.

The void problem, pointed out by J. Peebles \cite{PeeblesVoid} is hotly debated in the literature,
with arguments supporting his original claim \cite{TikhonovKlypin1,TikhonovKlypin2} as well as arguments to the contrary 
\cite{Patiri,Tinker1}.

The late merging problem appears in simulations which shows that accretion onto giant elliptical
galaxies at cluster centers continues until z = 0 \cite{Gao} while the independence
of the color-magnitude relation of SDSS\footnote{\textit{Sloan Digital Sky Survey} http://www.sdss.org/} galaxies on their environment \cite{Hogg2004} 
and the remarkable stability of the color-magnitude relation at z = 0.7 \cite{Cassata2007}
is consistent with the picture of giant galaxies as island universes, contrary to $\Lambda$CDM simulations.   
Likewise, according to J. Peebles, the very existence of large galaxies like the 
Milky Way, with its spiral structure intact, suggests a lack of major mergers in recent history (see \cite{Disksurv} for thin disk dominated galaxies survival issues).
There is observational evidence that the latest ``major invasion'' happened 
to our Galaxy 10-12 Gyr ago \cite{Gilmore2002}. There are
also arguments to the contrary, claiming that Milky Way-like galaxies survive late merging
in N-body simulations \cite{LCDMdisks}.  

For an excellent discussion 
of the observational situation and comparison with the $\Lambda$CDM model, see Refs. \cite{6puzzles,NGP,PeeblesLambda-DE}, 
and the references therein.

Theoretical suggestions of Peebles and Gubser were followed by the work of Nusser \textit{et al.} \cite{NGP},
exploring the cosmological implications with N-body simulations. Some preliminary results on similar scalar model were also 
obtained by Rodr{\'{\i}}guez-Meza \textit{et al.} \cite{Rodriguez-Meza1,Rodriguez-Meza2,Rodriguez-Curves}. The Long-Range Scalar Interaction (LRSI) model started a debate 
in the literature recently, mainly focused on the weak-equivalence principle violation and it's impact on dynamic of Milky-Way satellites \cite{Satellites1,Satellites2,Satellites3}.

The work, presented here should be regarded
as the next step in this line of research. Like Nusser \textit{et al.}, we study the two-point correlation
function and the statistical properties of the mass distribution of the dark matter halos. To our
knowledge, for the first time in the literature, we also study the power spectra for a set of 
scalar field parameters with comoving screening length. Analogous model was analyzed in great detail by
 Grawdohl \& Frieman nearly 20 years ago \cite{GradwohlFrieman1,GradwohlFrieman2}. However, their model assumed
a fixed physical screening length, which is a fundamental difference in comparison to our model.
We have used more particles in our simulations compared to those
used by Nusser \textit{et al.}, and our resolution is better than their resolution; we also consider a wider range
of scalar model parameters. As a consequence, we resolve the power spectrum near the screening
length characteristic scale. Our results show a clear feature in the power spectrum near
a wave number,that is the inverse of the screening length. The extra power seen at higher
wavenumbers is generated by the gravitational field, enhanced by the scalar interaction.
Nusser \textit{et al.} could not see a similar feature in their correlation function because the
screening lengths they considered were too close to the mean interparticle separation in their
simulations. Apart from this important difference, we confirm their results. The scalar
field generates lower density in voids. It also shifts structure formation to higher redshifts.

This paper is organized as follows:
In section \ref{model} we introduce the effective
gravitational potential and modified force law used as an approximation of the scalar field. In Section
\ref{sim} we describe our N-body simulations. Our results are presented in Section \ref{results}.
A brief summary and discussion appears in Section \ref{CR}.

\section{\label{model}Theory}

Following Refs. \cite{GP1,GP2}, we consider dark matter (DM) particles as strings. 
Their dynamics is defined by conventional gravity as well as 
an additional attractive force, induced by an exchange of a massless scalar. 
This force is well represented by a Yukawa-like potential with a characteristic screening length dynamically generated  by the presence of the light particles coupled to scalar (see Ref. \cite{FarrarPeebles}
 for details on the dynamical screening mechanism).
We consider one species of strings as DM particles.
The force between two DM particles, each of mass $m$, arises from the potential
\beq
\label{eqn:Yukawa}
\Phi({\bf r}) = -\,{Gm\over r}\,\,g(x)\,\,,
\eeq
with
\beq
\label{eqn:g(x)}
g(x) \,=\, 1 + \beta\, e^{-x/r_s} \, .
\eeq
Here $G$ is Newton's constant; ${\bf r}$ and ${\bf x} = {\bf r}/a(t)$ are, respectively, 
the particle separation vector in real space and comoving
coordinates; $t$ is the cosmological time; $a$ is the scale factor, normalized to unity at present,
\beq
a(t_0) \; = \; 1 \; .
\eeq
Here and below the subscript ``$0$'' denotes the present epoch.
The parameter $r_s$ is the screening length and $\beta$ is a measure of the relative strength of the 
scalar interaction compared to conventional gravity. The screening length $r_s$ is constant in comoving coordinates 
because of the dynamical screening mechanism, specific to the class of scalar fields considered here.
Accordingly, modified potential gives rise to the modified force law between DM particles. The new appropriate form is
\beq
F_{DM} = -G{m_1\cdot m_2\over r^2}\left[1+\beta\left(1+{r\over r_s}\right)e^{-r\over r_s}\right]
\label{eqn:force-law1}
\eeq
We can adopt this modification as a distance dependent correction term to ordinary Newton force law:
\beq
F_{DM} = F_{Newton}\cdot F_s(r,\beta,r_s),
\label{eqn:force-law2}
\eeq
where $F_s$ characterize deviations from standard gravity:
\beq
F_s(r,\beta,r_s) = 1+\beta\left(1+{r\over r_s}\right)e^{-r\over r_s}
\label{eqn:force-law3}
\eeq
For $\beta=0$ or $r\gg r_s$ we have $F_s\rightarrow 1$, thus we recover standard Newtonian force law.

Switching from the discrete particle picture to fluid dynamics, we will now introduce the dark matter density
field, given by the expression
\beq
\label{eqn:deltarho}
\rho({\bf x},t) = \left<\rho\right>\,(1 + \delta) \, ,
\eeq 
where $\left<\rho(t)\right>$ is
the ensemble average of the dark matter density at time $t$, and $\delta({\bf x},t)$ describes local deviations from homogeneity.
The structure formation is driven only by the spatially fluctuating part of the gravitational potential, $\phi({\bf x},t)$,
induced by the density fluctuation field $\delta$,
\beq
\label{eqn:scalar-pot-r} 
{\phi({\bf x})\over G{\left<\rho\right>} a^2 }\, =\, - \int {d^3{\bf x'}\delta({\bf x'})\over|{\bf x-x'}|}
g\left( |{\bf x-x'}| \right)\, .
\eeq
The Fourier transform of this equation is 
\beq
\label{eqn:scalar-pot-k-space} 
\phi_{\bf k} = - {3H_0^2\Omega_M\over 2a}\,{\delta_{\bf k}\over k^2}\,\left[1+{\beta\over 1+(kr_s)^{-2}}\right]\,,
\eeq
where 
\beq
\label{eqn:phiK}
\phi_{\bf k} \, \equiv \, (2\pi)^{-2/3}\,\int \phi({\bf x})\,e^{-i{\bf k \cdot x}}\,d^3{\bf x}
\eeq 
and 
\beq
\label{eqn:delK}
\delta_{\bf k} \, \equiv \, (2\pi)^{-2/3}\,\int \delta({\bf x})\,e^{-i{\bf k \cdot x}}\,d^3{\bf x}
\eeq
are the Fourier transforms of $\phi({\bf x})$ and $\delta({\bf x})$,
respectively; ${\bf k}$ is the comoving wavevector, and the quantities
\beq
\label{eqn:Omega}
\Omega_M \,\equiv \,8\pi G \left<\rho\right>_0\,/3H_0^2
\eeq
and $H_0$ are, respectively, the present values of the dimensionless mean dark matter density
and the Hubble parameter. From now on, we will also use the symbols $h$ and $\Omega_{\Lambda}$,
denoting the dimensionless
$H_0$, expressed in units of 100 km s$^{-1}$Mpc$^{-1}$, and the cosmological constant contribution to the present
mean density. 

Note that when $\beta = 0$ or $x \gg r_s$, equations (\ref{eqn:scalar-pot-r}) and
(\ref{eqn:scalar-pot-k-space}) become identical with conventional gravity \cite{1980Peebles}. 
Indeed, consider the fractional deviation from the Newtonian gravitational potential,
\beq
{\Delta\phi_{\bf k}\over \phi_{\bf k}} \; \equiv \; {\phi_{\bf k} - \phi_{\bf k}^{\rm N}\over\phi_{\bf k}^{\rm N}} \; .
\label{eqn:Delta_phi}
\eeq
Throughout this paper, the label 'Newton', and the subscript 'N'  refer to the $\Lambda$CDM cosmology with 
the conventional Newtonian gravity. Equation (\ref{eqn:scalar-pot-k-space}) gives
\beq
{\Delta\phi_{\bf k}\over\phi_{\bf k}} \; = \; {\beta\over 1+(kr_s)^{-2}} \;\;\; .
\label{eqn:PHIcorrection}
\eeq
This is the Fourier image of the spatial decline of the Yukawa potential: in the limit
$kr_s \rightarrow 0$, the above expression vanishes. In the opposite limit, 
the fractional deviation from Newtonian gravity reaches a finite value,
\beqa
\Delta \phi_{\bf k}/\phi_{\bf k} \rightarrow \beta \quad {\rm for}\;\; kr_s\gg 1 ,\\
\Delta \phi_{\bf k}/\phi_{\bf k} \rightarrow 0 \quad {\rm for}\;\; kr_s\ll 1 \; .
\label{eqn:potential_limit}
\eeqa
We consider values of $\beta$ of order unity and screening lengths of order of 
few megaparsecs or smaller.
Therefore, we can expect that our model predictions differ from the $\Lambda$CDM cosmology only on scales $\sim 1\hmpc$,
while on larger scales these two models are indistinguishable, unlike other modifications of gravity, considered recently,
for example the DGP model \cite{DGP},  $f(R)$ theories \cite{mod_grav} ,modifications of the Newtonian gravity on megaparsec scales \cite{Sealfon2005, Shirata1, Shirata2, altgrav_nsim} or MONDian cosmological simulations \cite{KnebeGibson}.

Here we study only the distribution of the dynamically dominant dark matter particles. We will study
the baryon distribution as well in our future work.

\section{\label{sim} Simulations}

In this section we describe our numerical experiments.

\subsection{Initial Conditions}

To set up the initial conditions, we have to define the power spectrum of the dark matter density fluctuations,
\beq
\label{eqn:P(k)}
P(k) \, = \, \left< |\delta_{\bf k}|^2\right> \, .
\eeq
We use a power spectrum,
derived from the \verb#cmbfast# code by Seljak \& Zaldarriaga \cite{cmbfast} with cosmological parameters $h = 0.7$,
$\Omega_M = 0.3$, $\Omega_{\Lambda} = 0.7$ and $\sigma_8 = 0.8$. The last in this set of parameters is the
present value of the root-mean-square density contrast of dark matter spatial fluctuations within a 8 ${\hmpc}$ sphere. 
This is the conventional normalization parameter and a measure of the degree of inhomogeneity of the dark matter 
distribution.
 
The resulting power spectrum, together with the 
\verb#PMcode# by Klypin \& Holtzman \cite{PMcode} is used to displace particles from their regular lattice 
positions, following the Zel'dovich approximation (see Ref. \cite{Efstathiou85} for details).
The number of individual simulations for each set of model parameters has to be large enough to allow a decent average over
simulation-to-simulation phase fluctuations. We decided that for our purposes ``a large enough number'' is 10 realizations for $200\hmpc$ box simulations. 
In this manner we have obtained an ensemble average of simulations in the large box.

\subsection{The codes}

We use  
the freely available codes: (1) \verb#AMIGA# (\textit{Adaptive Mesh Investigations of Galaxy Assembly}) by Knebe, Green, Gill \& Saar which is the successor of the \verb#MLAPM# code \cite{MLAPMpaper} 
and (2) \verb#GADGET2# by Volker Springel \cite{Gadget1,Gadget2}. \verb#AMIGA# is a \textit{Particle Mesh} code with implementation of \textit{the Adaptive Mesh Refinements}(AMR)
 technique to obtain high force resolution. \verb#GDAGET2# is a \textit{Tree-Particle Mesh} code. We use \verb#AMIGA#'s pure \textit{Particle Mesh} (PM) kernel for large box simulation, reducing the simulation
 time at the expense of the force resolution. For study of the halo clustering properties we used \verb#GDAGET2#. To accommodate for the poor mass resolution we have divided our simulations into two sets. 

The first set of simulations was used to study the power spectrum and the spatial correlation function of the dark matter
density fluctuations and density probability distribution functions. The simulation box in this series of simulations has a width $200\hmpc$, allowing
 proper treatment of the fundamental mode of density perturbations, which remains in the linear regime at 
redshift $z = 0$. These are pure PM simulations.
 
In the second set of simulations we have used a box of width $25\hmpc$. These simulations are used to study the cumulative halo mass function and the redshift evolution of halo abundances and $p(\delta)$. This approach provides a
better force and mass resolution, but due to the smallness of the box we lack some power in large scales. For a discussion of the influence of the box size on the dynamics and statistics
of simulations, see Ref. \cite{haloes_box,boxsize}.

In each experiment we save the particle positions and velocities at redshifts
5; 3; 2; 1; 0.9; 0.8; 0.7; 0.6; 0.5; 0.4; 0.3; 0.2; 0.15; 0.1; 0.05, and the redshift of the final output, $z = 0$.
This archive is used to study the evolution of the dark matter distribution with redshift.
For an experiment, involving $N_p$ dark matter particles in a box of size $L$ at present, 
the particle mass, $m_p$, is given by
\beq
m_p \, = \,\left<\rho\right>\,(L^3/N_p).
\eeq
For our simulations, $N_p = 128^3$ for 200 $\hmpc$ box and $N_p = 256^3$ for 25 $\hmpc$. The other important simulation parameters are the force
resolution $\varepsilon$, the interparticle separation, 
\beq
\ell \; = \; (L^3/N_p)^{1/3} \;\; , 
\eeq
and the Nyquist wavenumber,
\beq
k_{\rm Nyq} \; = \;\pi/\ell \;\; .
\eeq 
\begin{table*}
\caption{\label{tab:sim_params} Simulation parameters. 
$L \,$ is the box size $[\hmpc]$; $\, z \,$ is the initial redshift; 
$\, m_p \,$ is the mass of a single particle $[h^{-1}M_{\odot}]$; $\, \varepsilon \,$ is the force resolution $[\hkpc]$; $\, \ell \,$ is
the mean interparticle separation $[\hmpc]$; $\, k_{\rm Nyq} \,$ is the Nyquist wave number $[h{\rm Mpc}^{-1}]$; 
$\, r_s \,$ is the screening length $[\hmpc]$; and $\, \beta \,$ is the relative strength of the scalar force.}
\begin{ruledtabular}
\begin{tabular}{lllllllll}
$L$ & $\;z$  & $ m_p$ & $ \varepsilon$ & $\ell$ 
& $k_{\rm Nyq}$ & $ r_s$ & $\beta$\\
\hline 
200 & 30  & $3.18\cdot 10^{11}$ & 800 & 1.563 & 2.01 & $1; 2; 5$ & $-0.5; 0; 0.2; 0.5; 0.7; 1$\\
\hline
25 & 60 & $7.7\cdot 10^{7}$ & 10 & 0.097 & 32.1 & $0.5; 1; 1.5; 2$ & $0; 0.2; 0.5; 1$ 
\end{tabular}
\end{ruledtabular}
\end{table*}

We list the above parameters, evaluated for the large and the small box, in Table \ref{tab:sim_params}.

\subsection{The Green's Function}

For conventional gravity in Fourier space, the discrete Poisson equation can be written as
\beq
\phi_{\bf k} \,= \, {3H_0^2\Omega_M\over 2a}\,\delta_{\bf k}\,{\mathcal G}_{\bf k}\, ,
\label{eqn:poisson_f_std}
\eeq
where ${\mathcal G}_{\bf k}$ is the Green's function. We use the seven-point finite-difference approximation to the
Green's function to solve the Poisson's equation in a cubic box
with periodic boundary conditions. It is defined for a discrete set of arguments, 
\beq
\label{eqn:q}
{\bf q} \,= \,\{q_j\} \,=\, {\bf k}\cdot L/(2\pi) \, ,
\eeq
where $L$ is the present size of the comoving simulation box. The subscripts
$j = 1,2$, or $3$ denote the three dimensions in ${\bf k}$ space. The number of grid cells in each dimension
is ${\mathcal N}$ (256 in our simulations), so the cell numbers assume integer values in the
range
\beq
\label{eqn:1...N}
q_j \,=\,1,2,\, \ldots\,, {\mathcal N} \, .
\eeq
Each integer triple defines the grid point, at which the Green's function is evaluated:  
\beq
{\mathcal G}_{\bf q}^{\rm N}  = -\pi \left[ {\mathcal N}^2 \sum_{j=1}^3 \sin^2 (\pi q_j / {\mathcal N})\right]^{-1} \,.
\label{eq:green_std}
\eeq
Using the equation. (\ref{eqn:scalar-pot-k-space}), we modify Green's function 
to get the proper potential for our model of scalar interactions:
\beq
{\mathcal G}_{\bf k}^{\rm scalar} = {\mathcal G}_{\bf k}^{\rm N}\,\,\left(1+{\beta\over 1 + (kr_s)^{-2}}\right) \,\,.
\label{eq:green_mod}
\eeq
The \verb#AMIGA# code and PM part of the \verb#GADGET2# code uses equation (\ref{eqn:poisson_f_std}) and the fast Fourier transform
technique to evaluate the gravitational potential at grid points. \verb#GADGET2# use Oct-Tree algorithm to calculate short distance 
part of the particle forces, we have modified this part of the code using equations (\ref{eqn:force-law2},\ref{eqn:force-law3}) accordingly.
Then particle positions and momenta are updated in the standard way for N-body algorithms.

\subsection{Particles and Halos}

To identify collapsed objects from now on called  'halos' we have used 
\verb#AMIGA#'s Halo Finder (\verb#AHF#) by Knollman \& Knebe \cite{AHF} which is 
the successor of the \verb#MHF#  halo finder by Gill, Knebe \& Gibson \cite{MHFpaper}. 
The \verb#AHF# uses AMR to find halo centers, then it probes the halo density
profile around each center in nested radial bins until
the spatially averaged density contrast reaches the virial overdensity, $\, \Delta$.
At $z=0$, in a $\Lambda$CDM universe \cite{GrossThesis}, $\, \Delta =
340$. Given our mass resolution in the bigger box, we can expect to follow the assembly of 
halos, corresponding to clusters and superclusters, and study the statistics of clustering at large scales. 
Simulations in the small box should be good enough to investigate the assembly of much less massive objects, 
like halos of clusters and galaxies.

It is important to bear in mind that at small scales we are
limited by force resolution. The smallest size of
gravitationally bound objects that can form in our box is $2\varepsilon$  (see Table \ref{tab:sim_params} and 
Ref. \cite{Knebe_resolution}). We also suffer from the well-known problem of \textit{overmerging} described in great detail by Klypin \textit{et al.} in Ref. \cite{overmerging}. As a consequence, 
our simulations underestimate low-mass object abundances.

\section{\label{results}Results}

In this section we present the results obtained in our numerical experiments. 
We provide maps of the spatial distribution of dark matter particles
as well as clumps of particles, called halos. We also study different statistical 
measures of clustering, such as the
power spectrum, the two-point correlation function, density probability distribution function, and the cumulative halo mass function with and
without the scalar interaction.

\begin{figure*}
\resizebox{182mm}{!}{\includegraphics[angle=-90]{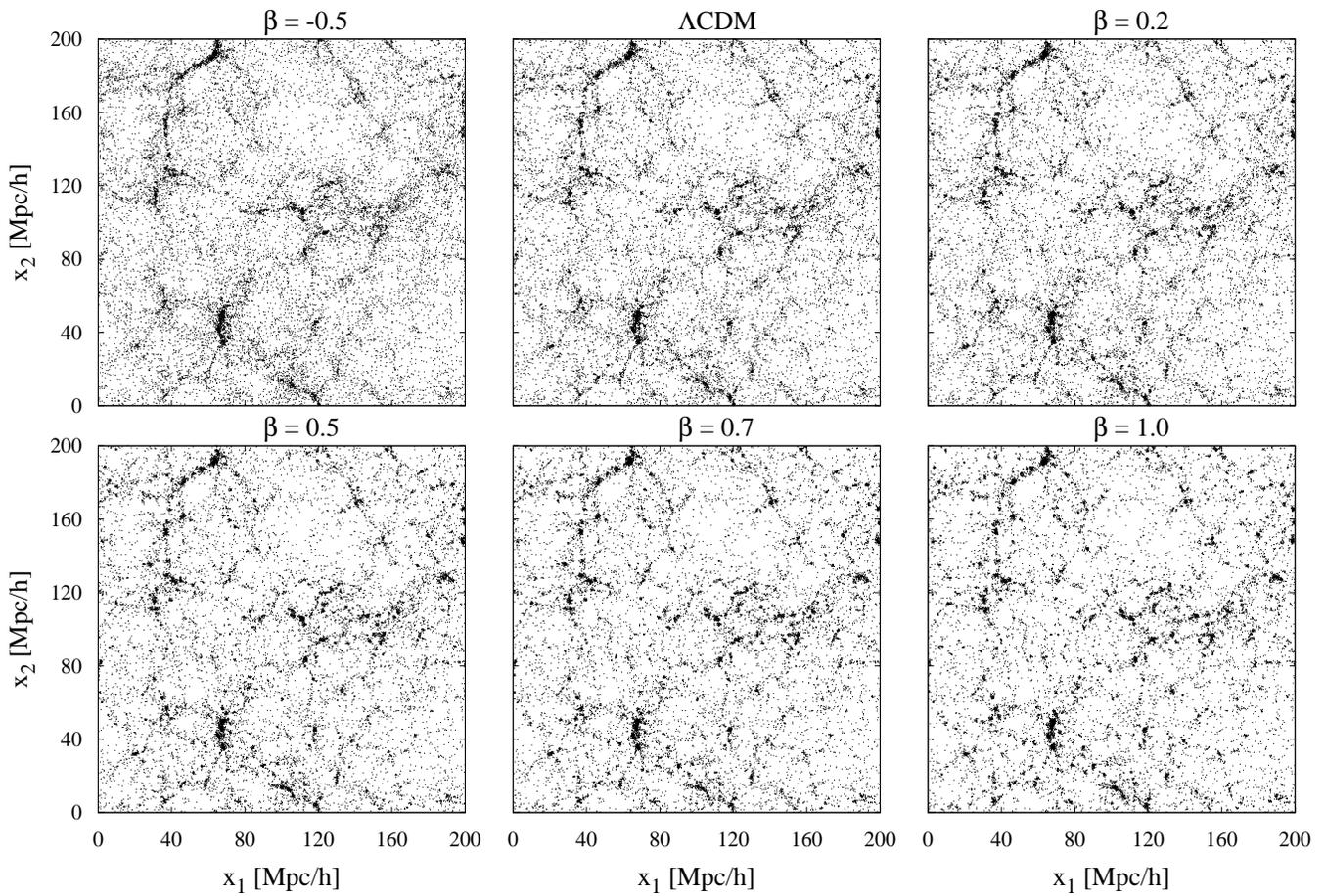}}
\caption{Slices cut through centers of simulation boxes of width $200\hmpc$ at $z = 0$. 
The plots show particle positions, ${\bf x}$, in the $(x_1,x_2)$ plane. 
Each slice has dimensions $200\times200 (\hmpc)^2$ in this plane and
a thickness of  $10\hmpc$ along the $x_3$ coordinate axis. 
\label{fig:dens200}}
\end{figure*}

\subsection{\label{pattern}The Clustering Pattern}

Figure \ref{fig:dens200} shows the final particle distribution in $10\hmpc$ slices,
cut through the centers of $200\hmpc$ simulation boxes. For clarity each slice shows only 1/10 of the total
number of particles, selected at random \footnote{Random seed number is the same for each slice, so we compare particles with the same ID's in each simulation box.} 

The frame, labeled '$\lcdm$' in Figure \ref{fig:dens200} 
assumes $\beta = 0$, while the remaining frames show particle distributions for a set of different $\beta$ parameters and a fixed screening length,  $r_s = 2\hmpc$. 
As a test, we also consider 'antigravity' with $\beta = -0.5$. All of the frames in Figure \ref{fig:dens200} have evolved from 
the same initial state, with identical amplitudes and phases of density fluctuations, set up at $z = 30$. 

\begin{figure*}
\resizebox{120mm}{!}{\includegraphics[angle=-90]{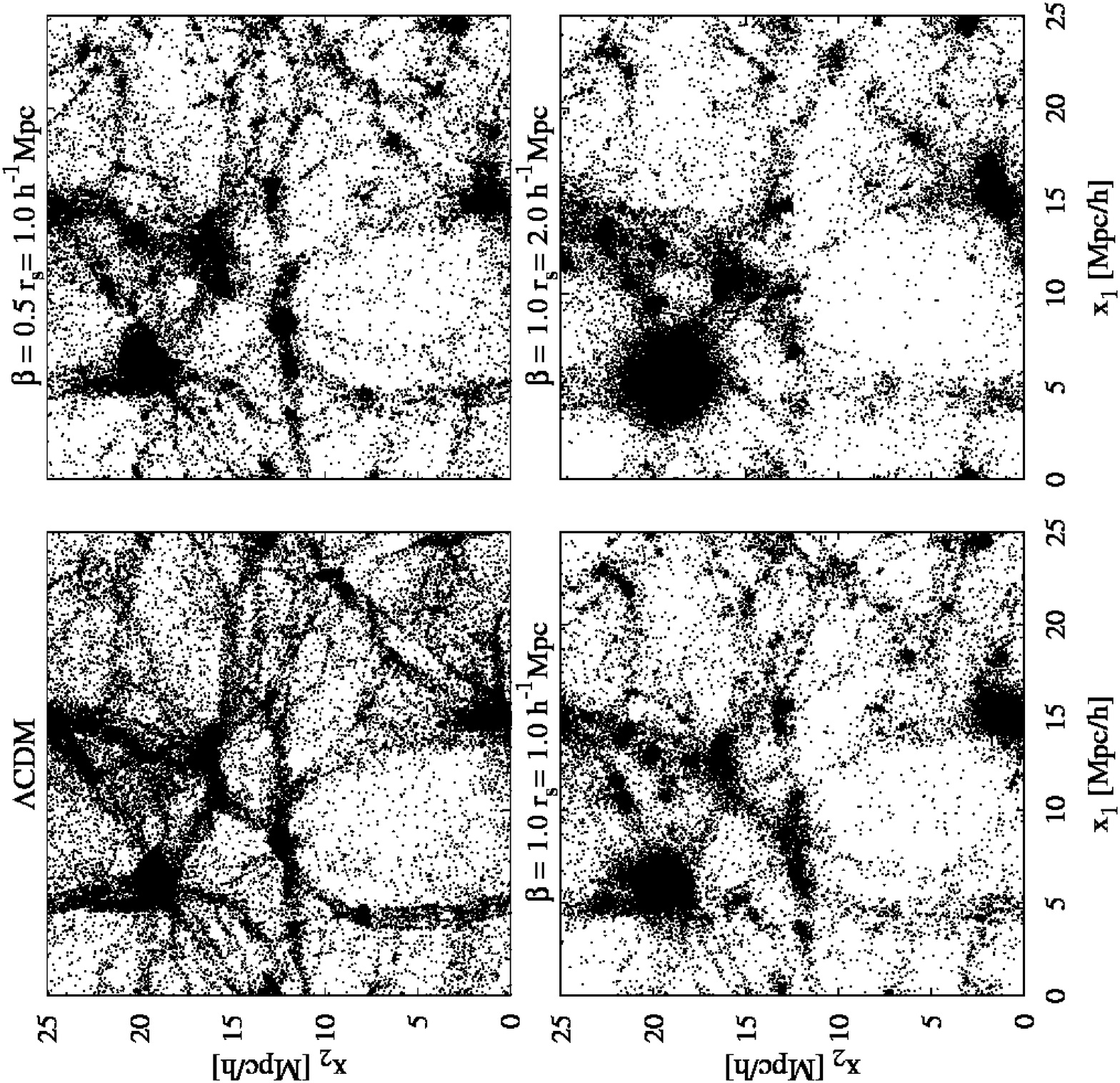}}
\caption{
Similar results for the smaller box, $L = 25\hmpc$. The thickness of each slice is 
$1.5\hmpc$. 
\label{fig:dens25}}
\end{figure*}

At first sight, modified gravity simulations look similar to the $\lcdm$ case. The filaments, voids and high-density peaks occupy 
similar positions in the frames. However, close inspection shows that with increasing $\beta$, the 
voids appear increasingly more empty.
This phenomenon is seen more clearly in Figure \ref{fig:dens25}, where we show $1.5\hmpc$ slices, cut through 
the centers of our smaller simulation boxes ($L = 25\hmpc$). 

Figure \ref{fig:dens25} also shows that the scalar forces enhance the
accretion of matter, producing more massive halos at small scales. 
Pancake-like structures, seen in the $\beta = 0$ frame, are more homogeneous
than their counterparts in frames with $\beta > 0$, which show substructure.
The fragmentation into subsystems of smaller halos is increasingly more pronounced,
with increasing values of $\beta$. Quantitatively, 
we can expect an increase of the amplitude of the power spectrum at small scales, corresponding to wave numbers
$k \gtrsim r_s^{-1}$, and an opposite effect for $\beta = - 0.5$, when the pancakes appear even
more homogeneous than those generated by Newtonian gravity.

\subsection{The Power Spectrum}

The power spectrum is a convenient measure of the strength of dark matter clustering. It is well constrained by 
redshift surveys of galaxies, such as the SDSS
catalog. In the longwave tail, 
corresponding to wave numbers 
\beq
\label{eqn:SDSSrange}
0.01 \,{\rm Mpc}^{-1}h \; \leq k \; \leq \; 0.3 \,{\rm Mpc}^{-1}h \;\; , 
\eeq
this survey provides a reliable estimate of $P(k)$ \cite{SDSS2006paper}. In this range,
the  $\Lambda$CDM power spectrum agrees well
with observations. We also use non-inear fit to the power spectrum presented by Smith \textit{et al.} in \cite{Smith_pk} as 
a measure of theoretical $\lcdm$ P(k). This can be used to constrain scalar field models. 

\begin{figure}
\resizebox{86mm}{!}{\includegraphics[angle=-90]{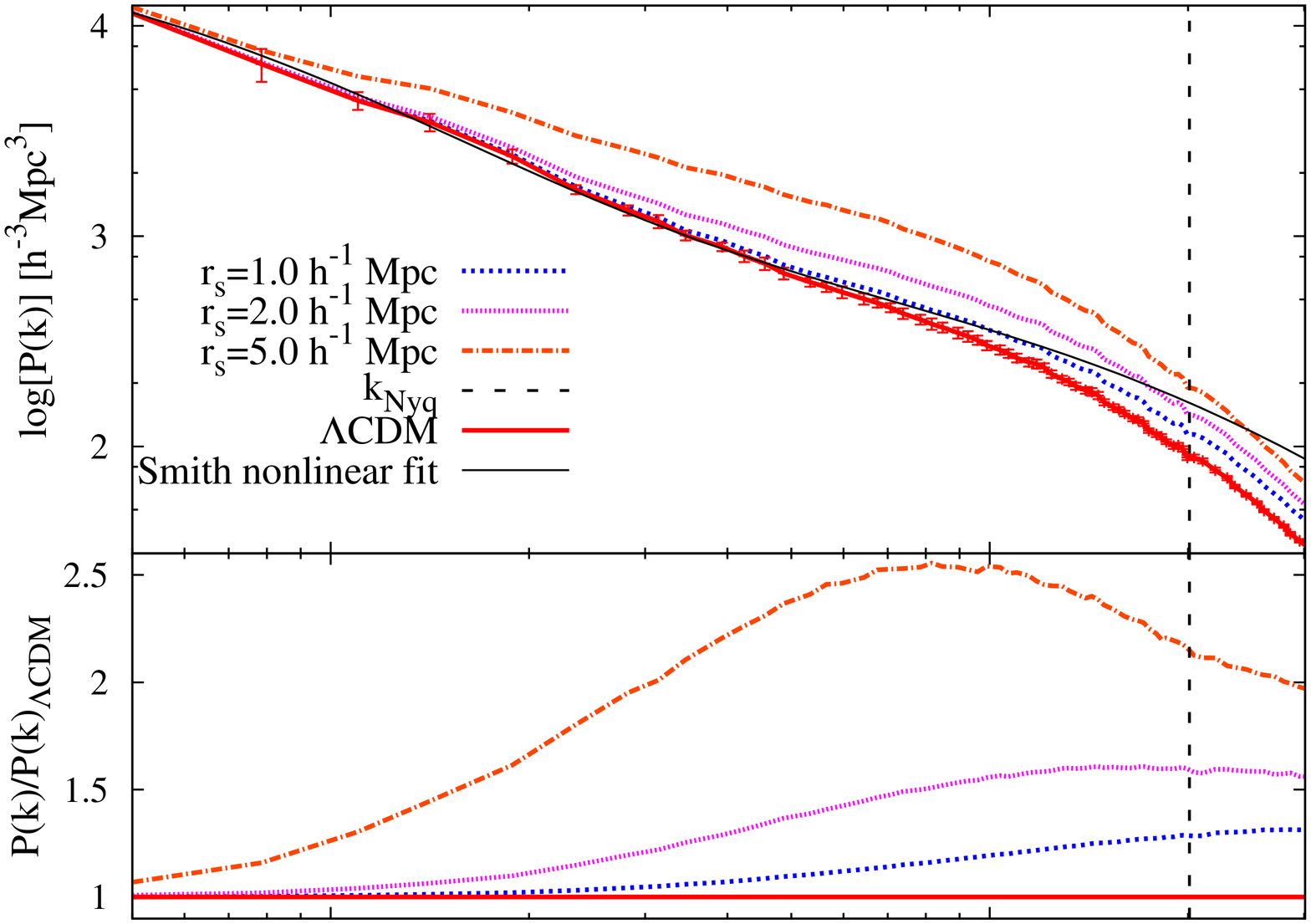}}\\
\resizebox{86mm}{!}{\includegraphics[angle=-90]{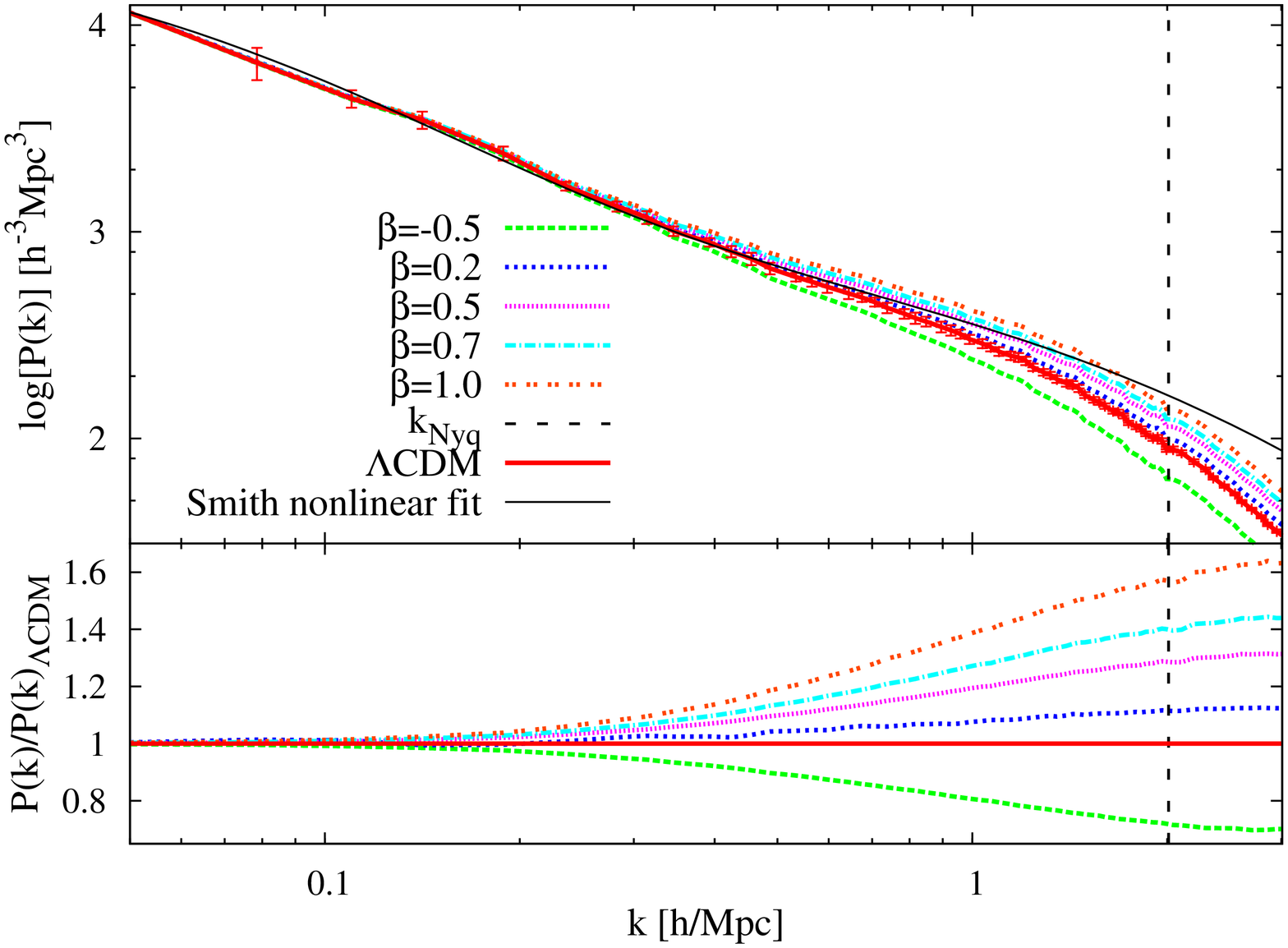}}
\caption{Power spectra from simulations with $L = 200\hmpc$. Vertical dashed lines show the Nyquist 
wave number, while the thin black line is the nonlinear fit for $\lcdm$ $P(k)$ from Smith \textit{et al.} We also plot the ratio of the scalar interaction induced $P(k)$ to its Newtonian
counterpart, $P_{\rm N} (k)$. For all of the upper pair of plots  $\,\beta = 1$, while the
$\, r_s\,$ parameter is allowed to vary. For the bottom pair $\, r_s = 1\hmpc\,$ is fixed,
while the value of $\,\beta \,$ changes from $\,-0.5\,$ to $\,1$.
Each curve was obtained by averaging over 10 realizations with different initial phases.
\label{fig:pk_comp200}}
\end{figure}
\begin{figure}
\resizebox{86mm}{!}{\includegraphics[angle=-90]{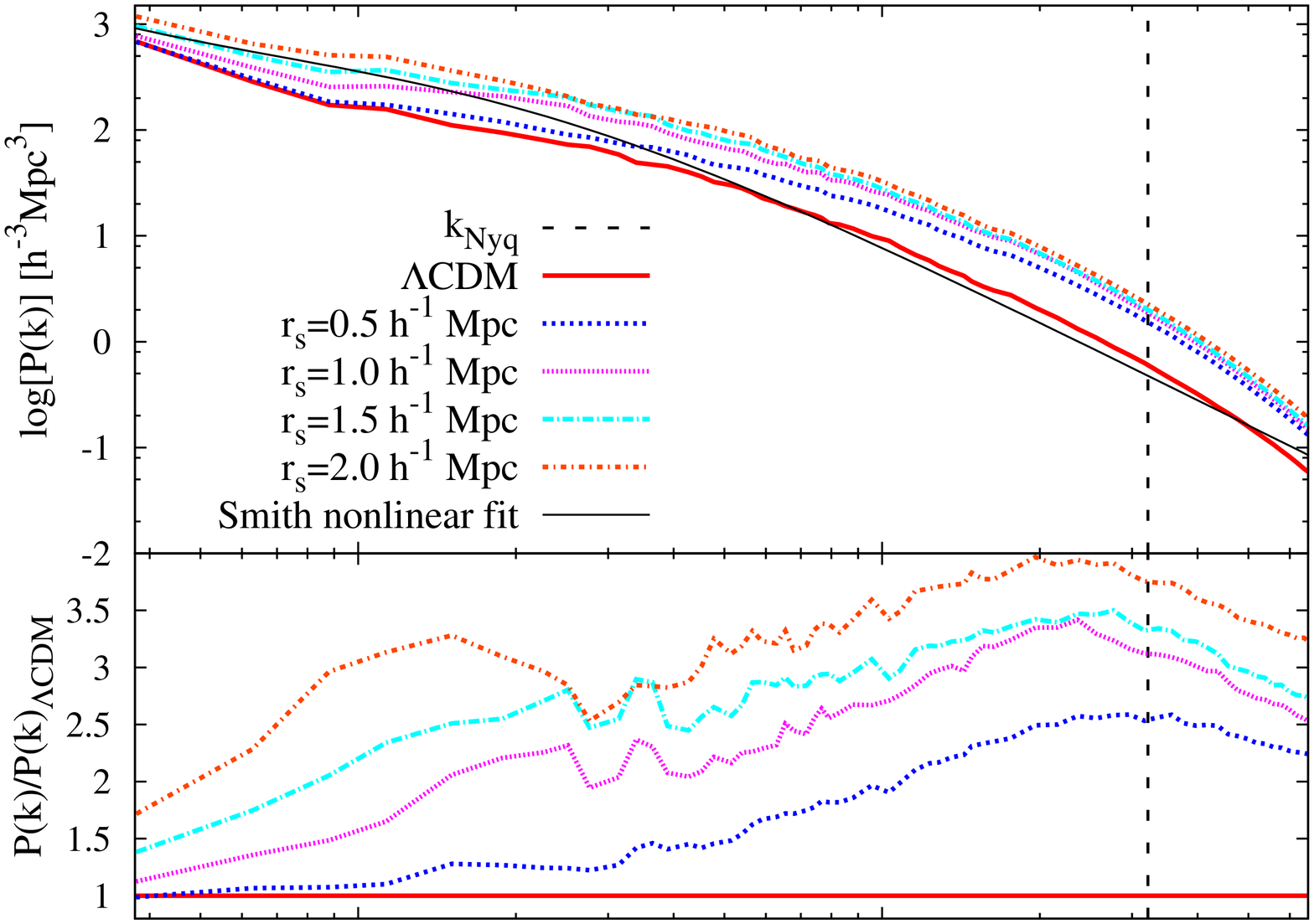}}\\
\resizebox{86mm}{!}{\includegraphics[angle=-90]{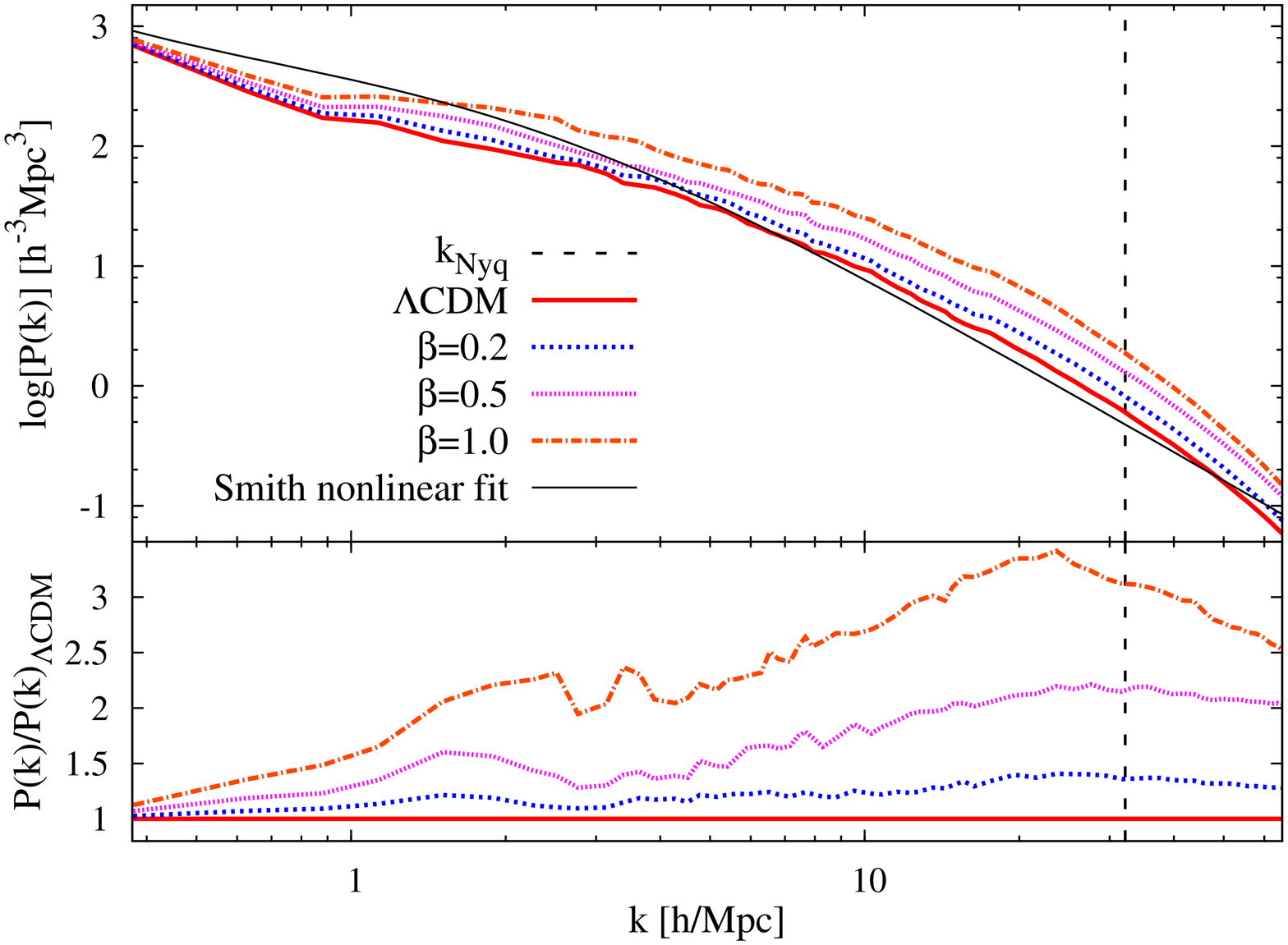}}
\caption{Power spectra plots similar to those in figure \ref{fig:pk_comp200},
obtained from a set of simulations, using the smaller box, $L = 25\hmpc$.  
\label{fig:pk_comp25}}
\end{figure}

In Figure \ref{fig:pk_comp200} we plot the final power spectra, obtained from the simulations with
$L = 200\hmpc$. The top pair of plots shows $P(k)$ and the ratio $P/P_{\rm N}$ for scalar forces 
with fixed $\beta = 0.5$ and a varying
screening length. The pair of plots  below was obtained from simulations with
a fixed screening length, $r_s = 1\hmpc$, and a varying $\beta$. 

These plots show that the rate of growth of density perturbations is more sensitive to the 
range of the scalar field, $r_s$, than to its strength, as long as $\beta > 0$.
The scalar force increases the rate of growth of density fluctuations on comoving scales, smaller
than $r_s$ and wave numbers $k \gtrsim r_s^{-1}$. As expected, there is an opposite effect for $\beta = - 0.5$:
the rate of growth of density perturbations at small scales is reduced.

We describe the deviations from the Newtonian power spectrum at a given wave number by the parameter
\beq
\label{eqn:deltaP}
{\Delta P\over P}\,\equiv \, {P - P_{\rm N}\over P_{\rm N}} \;\;\; .
\eeq
These deviations remain negligible on large scales for all of the considered models with one
exception: the model with $r_s = 5\hmpc$. For this model, $\Delta P/P$ is significant on  
all scales present in the simulation box, including the SDSS wave number range.
Therefore all scalar field models with 
$r_s \geq 5\hmpc$ do not appear promising. 
To assign statistical significance to their failure to reproduce the real Universe, it is necessary to 
create a mock SDSS survey and reproduce the power spectrum estimator, used by the observers. 
We are planning to run appropriate Monte Carlo simulations and 
address this problem in greater detail in a forthcoming paper.

The model with $\,r_s = 5\hmpc\,$ may not be successful in reproducing the real Universe,
but it is extremely useful in understanding the physics of the scalar field because this
value of the screening length is much larger than the interparticle separation in the
big box, $\, \ell \approx 1.6 \hmpc$. Therefore, we can resolve the difference between
purely Newtonian gravity and the scalar field. Indeed, for $\, k = 1/r_s = 0.2\,$, 
and $\, \beta = 0.5\,$, we get $\, \Delta P/P = 0.5\,$ (see Figure \ref{fig:pk_comp200}),
and an increasing $\, \Delta P/P\,$ for larger wave numbers. The
gravitational attraction, enhanced by the scalar field generates the extra
power in the plot. To look for a similar jump in amplitude
for smaller values of $\,r_s\,$, we need a better resolution. So, we have to turn to the
simulations in the smaller box, where $\, \ell = 125 \hkpc$. And we do find a similar
feature: for $\,\beta = 0.5\,$ and $\, r_s = 1\hmpc\,$, we get $\, \Delta P/P = 0.5\,$
at $\, k = 1/r_s = 1 h{\rm Mpc}^{-1}\,$ (see Figure \ref{fig:pk_comp25})!

The decline of $P_{\rm N}/P$ with decreasing wave number in all considered models
is a natural consequence of the presence of the Yukawa cutoff, reflected in the 
equation (\ref{eqn:PHIcorrection}). 

In the opposite limit, with growing wave numbers, all power spectra in Figure \ref{fig:pk_comp200}
seem to misbehave. 
Equation (\ref{eqn:potential_limit}) suggests that all $P_{\rm N}/P$ curves should flatten
for large wave numbers, $\, k \gg 1/r_s$. Instead,
we see a decline. Since this behavior is in disagreement with gravitational dynamics, and 
since it appears in the range $k \gtrsim k_{\rm Nyq}$, we can expect that the 
decline is an artifact of the discrete nature of the simulation.
If this is indeed the case, then in the simulation in the smaller box, 
this artifact should move to higher wave numbers. This simulation 
has a smaller interparticle separation and a better force resolution.
Hence, for wave numbers in the range 
\beq
r_s^{-1} \;\lesssim \;k\; \lesssim \; k_{\rm Nyq} \;\; , 
\eeq
we should see a flat section
of the $P/P_{\rm N}$ curve, followed by a decline for $k \gtrsim k_{\rm Nyq}\;$.
This is exactly what happens in Figure \ref{fig:pk_comp25}.  
The $P_{\rm N}/P$ ratio rises with $k$ until it reaches its maximum,
followed by a plateau, and then decline for wave numbers above the Nyquist limit.

\subsection{\label{xi}The Correlation Function} 

Another convenient measure of the strength of clustering is the spatial two-point correlation function,
\beq
\xi\,\left( |{\bf x - x'}|,\,t\right)  \; = \; \left\langle \,\delta({\bf x},t)\delta({\bf x'},t)\,\right\rangle \; .
\label{eqn:xiDef}
\eeq
It is related to the power spectrum by the Fourier transform,
\beq
\xi\,(x) \; = \; (2\pi)^{-3}\,\int\,P(k)\,e^{i{\bf k \cdot x}}\, d^3{\bf k} \; .
\label{eqn:xi-P}
\eeq
Under ideal conditions, when the power spectrum is determined in the entire wave number range from zero to
infinity, $P$ and $\xi$ are equivalent to each other. However, measurements from simulations or galaxy
redshift surveys provide only noisy estimates of $P(k)$ or $\xi(x)$ for limited ranges of $k$
and $x$. Therefore, in numerical experiments, it is safer to estimate $\xi(x)$ directly from
the particle positions in the simulation output, using the expression (\ref{eqn:xiDef}), or
its discrete version, based on the probability density of finding a pair of particles in a
separation range from $x$ to $x+dx$ (see Ref. \cite{1980Peebles}).
The results presented in this section are therefore not a mere Fourier transform of the
results discussed in the two previous subsections.  
Following the approach we used to study power spectrum deviations from the Newtonian case, here
we introduce a similar measure of the deviations of the scalar-induced correlation function from its 
Newtonian cousin, $\xi_{\rm N}$. This is the quantity
\beq
\dfrac{\Delta \xi}{\xi} \;\; \equiv \;\; \dfrac{\xi - \xi_{\rm N}}{\xi_{\rm N}} \; .
\eeq
\begin{figure*}
\resizebox{188mm}{!}{\includegraphics[angle=-90]{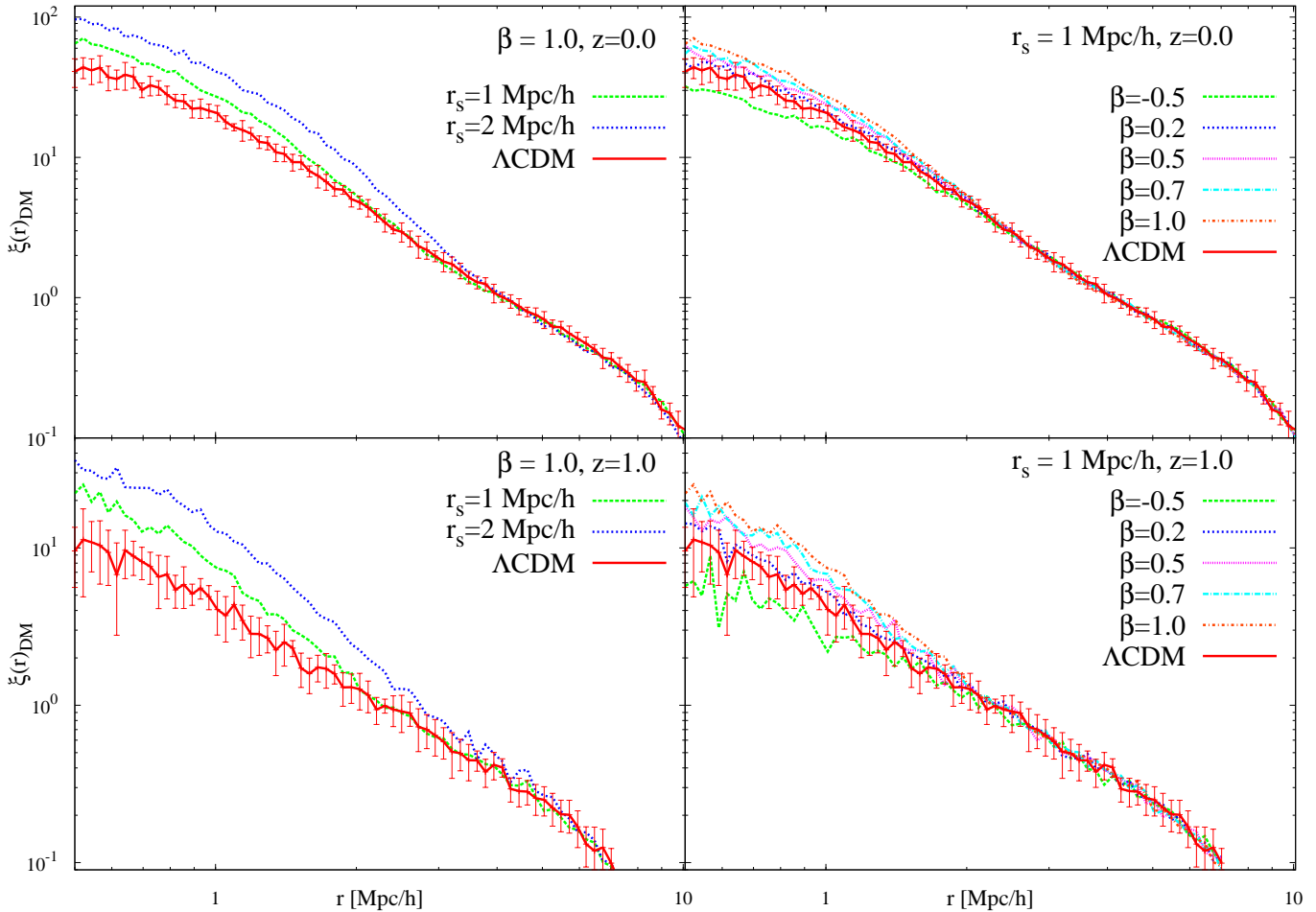}}
\caption{The correlation function for pairs of DM particles at redshift z = 1 (bottom frames) and z = 0 (top frames).
The width of the simulation box is $200\hmpc$. All plots were averaged over 10 simulations. 
To avoid overcrowding, we show 1 standard deviation error bars
for the Newtonian $\,\xi(x)\,$ only. The error bars for the scalar-induced correlation functions are similar.
Their size is determined mostly by the number of realizations and by the size
of the box. Both these quantities are identical for all simulations presented here.
\label{fig:xi_comp}}
\end{figure*}
The four frames in Figure \ref{fig:xi_comp} show two-point correlation functions for DM particles in the $200\hmpc$ box. 
The top two frames show simulation outputs at $z = 0$, while the bottom frames refer to an earlier epoch, $z = 1$. 
For clarity, we show one standard deviation errors only for the Newtonian case. 
The error bars for the remaining models are similar.

The large amplitude in  $\,\Delta \xi/\xi\,$ at separations $\, x \lesssim 1\hmpc$ 
is probably an artifact because the separations involved are smaller than $\,\ell = 1.56 \hmpc$.
Nusser \textit{et al.} \cite{NGP} have discovered a similar shoulder in $\, \xi(x) \,$ in
their simulations and they provided an interpretation similar to ours.
They have also pointed out that the correlation
function in their simulations does not possess a feature at $\, x = r_s\,$ ``despite the scalar
force attraction at smaller scales.`` We believe that the source of this problem is
poor resolution, not dynamics. The range of screening lengths they consider is
uncomfortably close to their interparticle separation. Our correlation functions, plotted
in Figure \ref{fig:xi_comp}, suffer from the same problem. As we have shown in our
discussion of the properties of simulated power spectra, this problem can be avoided
by improving the resolution. For an appropriate choice of the box size and the screening length,
when $\, r_s > 1/k_{\rm Nyq}\,$, the feature at $\,k = 1/r_s\,$ can be resolved.  
We therefore have no doubt - the feature is there. To rediscover for $\,\xi(x)\,$ what
we have already seen for $\,P(k)\,$, we only need
higher resolution simulations, like those in Ref.\cite{Millenium_run},
but with a modified Green's function.
\begin{figure*}
\resizebox{187mm}{!}
{\includegraphics[angle=-90]{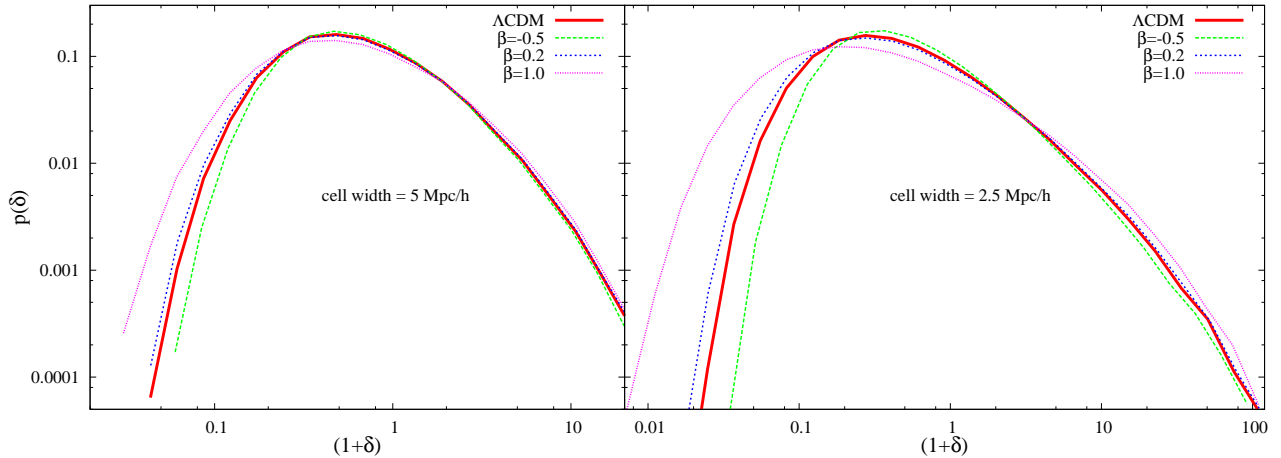}}
\caption{The density probability distribution functions obtained from $200\hmpc$ boxes. Results for fixed $r_s = 1
\hmpc$. Left panel shows $p(\delta)$ calculated using the grid cells width of $5\hmpc$, while in the right panel are functions calclated with cells width of $2.5\hmpc$.
\label{fig:dpdf1}}
\end{figure*}

Despite the modest resolution of the present results, we believe that qualitatively, 
the stratification of the correlation functions
with respect to $r_s$ and $\beta$, seen in Figure \ref{fig:xi_comp} reflect the true dynamics. 
Note that for $\, \beta <0 \,$, when gravity is weaker than Newtonian, we get $\,\Delta\xi/\xi\ < 0 \,$,
which is dynamically reasonable. 
 
Our results also differ from those of Nusser \textit{et al.} at larger separations,
$x > 3\hmpc$, where according to their Figure 7, $\, \Delta\xi/\xi < 0$.
This does not happen in our simulations unless $\beta < 0$. 
This discrepancy, however, becomes statistically insignificant
if we use our error bars for guidance (Nusser \textit{et al.} 
do not provide error bars for their plots). 

As we have already mentioned, for separations $x > r_s$, $\Delta\xi/\xi$
is consistent with zero for all models considered here. This is good news,
since the observed $\xi(x)$ at large separations agrees well
with the $\Lambda$CDM model.

Last but not least, it is worth noticing that in Figure \ref{fig:xi_comp},
$\Delta\xi/\xi$ is higher at $z = 1$ than at $z = 0$. 
The growth of Newtonian $\xi$ is retarded with respect to the
scalar-induced $\xi$ at high redshift. Later, 
$\xi_{\rm N}$ 'catches up' with the scalar $\xi$. As a consequence, $\Delta\xi/\xi$ is reduced. 
A possible explanation for this phenomenon is that 
in the presence of scalar forces, rapid matter accretion and formation of virialized objects occurs at earlier epochs 
than in the standard gravity model.  This interpretation is supported by our direct analysis of the process of halo
formation, presented below.

\subsection{\label{dpdf}Emptiness of the voids} 

\begin{figure*}
\resizebox{187mm}{!}
{\includegraphics[angle=-90]{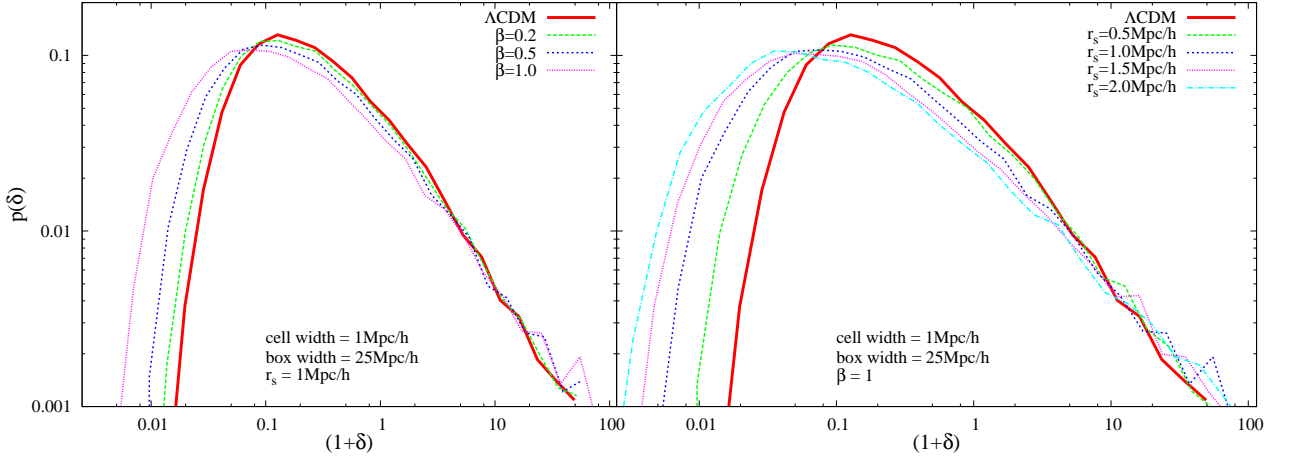}}
\caption{This figure is analogous tho the previous one. Here we present the $p(\delta)$ obtained from $25\hmpc$ boxes. The cell width used for calculation was $1\hmpc$. LRSI models in the left panel have fixed $r_s = 1\hmpc$, on the right panles $r_s$ vary and we keep $\beta = 1$ fixed.
\label{fig:dpdf2}}
\end{figure*}

As a measure of the emptiness of the voids in our simulations we will use the density contrast probability distribution function $p(\delta)$. To calculate these functions we extrapolate
particle positions to the uniform lattice grid to obtain density. After conversion to the density contrast we compute $p(\delta)$ numerically in the usual way, by using
\beq
dp(d\delta) = {\sum_{i}^{N_c}[\delta_{Dirac}(\delta_i = d\delta)]\over N_c}\;\;\;,
\label{egn:pdf_def}
\eeq
where the sum runs for all the cells, and $N_c$ is the total number of cells and $\delta_i$ is a density contrast in a $i$-th cell.
We have calculated the density contrast probability distribution function for the large and small boxes. The results are presented in figures \ref{fig:dpdf1} and \ref{fig:dpdf2}.

In Figure \ref{fig:dpdf1} we show $p(\delta)$ calculated in the $200\hmpc$ box. Functions in the left frame was calculated using the $5\hmpc$ grid cell width, while on the right
 frame we show density probability distribution functions obtained with the $2.5\hmpc$ grid cell width. In both cases we show the effect of varying the $\beta$ parameter keeping the screening length $r_s = 1\hmpc$. 
In Figure \ref{fig:dpdf2} we plot probability distribution functions (PDFs) oobtained from simulations in the small box. The density field was calculated on a uniform grid of cells with length of $1 h^{-1} Mpc$. On the left frame we show functions in
 models with fixed $r_s=1\hmpc$ and varying $\beta$. In the right frame we plot the results for models with fixed $\beta=1$ and various $r_s$ values.

Clearly we see, that introduction of LRSI stretches the $p(\delta)$ toward a smaller density contrast, while keeping the relative value at high density tail close to the $\lcdm$ case. 
This is striking confirmation of what was expected. LRSI produce emptier voids. The smaller the considered cell width the more pronounced the effect. 
An interesting effect seen in our distribution functions is the reduced skewness. This is different from standard gravity which increases the skewness \cite{RomanSkewness}.
We will study this effect in the future.

\subsection{Halos formation in the Small Box}
\begin{figure*}
\resizebox{125mm}{!}{\includegraphics[angle=-90]{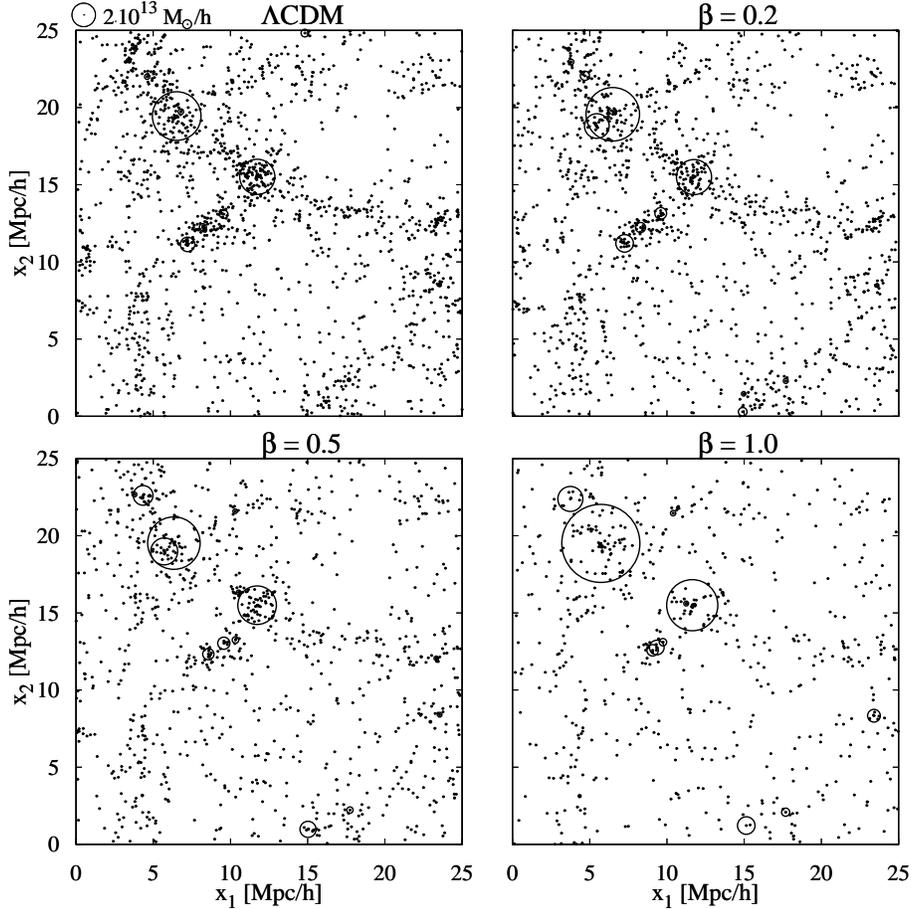}}
\caption{Positions of halo centers obtained from a particular simulation output. The size of the box is
$\, L = 25\hmpc$. To avoid overcrowding, we show only the halos with centers in 
the range $\, 0\leq x_3\leq 5\hmpc$. Halos with masses greater than $\, 5\cdot 10^{12}M_{\odot}\,$
appear as circles. Their diameters are proportional to their masses. Less massive halos are represented by dots. 
The top left frame shows the standard gravity case. The halo distribution in all of the remaining frames is plotted for
$\,r_s = 1\hmpc \,$ and three different values of $\, \beta$. The circle above the upper frame represents a halo with a mass of $\, 2\cdot 10^{13}M_{\odot}$.   
\label{fig:halos25}}
\end{figure*}
\begin{figure*}
\resizebox{125mm}{!}{\includegraphics[angle=-90]{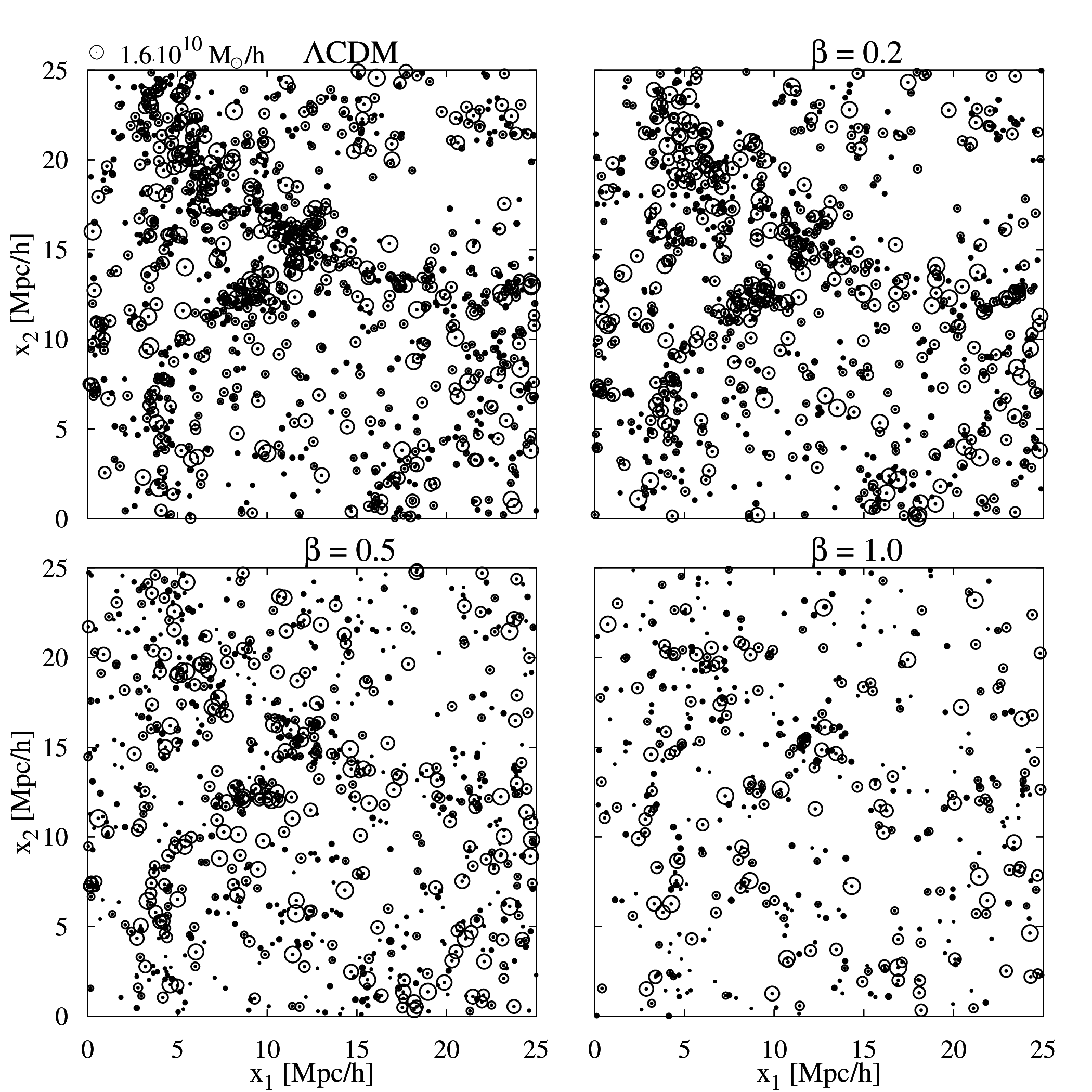}}
\caption{Maps of the halo distribution, derived from the same simulations as those shown 
in Figure \ref{fig:halos25}, but for halos with masses, smaller than ${\tt M} = 2\cdot10^{10}h^{-1}M_{\odot}$. 
Even smaller halos with masses $\,M \leq {\tt m} = 8\cdot10^{9}h^{-1}M_{\odot}$ are represented by dots.
Those with masses in the range $\, {\tt m} < M \leq {\tt M}\,$ appear as circles.
\label{fig:lm_halos25}}
\end{figure*}

In this section we study the impact of modified gravity on the halo formation process.
To identify halos, we apply the halo finder program (the \verb#AHF#, introduced earlier) to output 
snapshots from the $25\hmpc$ simulations. We study 
all halos with more than 20 particles, 
hence our minimal halo mass is $1.5\cdot 10^{9}h^{-1}M_{\odot}$. 

\subsubsection{High Mass Halos}

In Figure \ref{fig:halos25}
we plot the present comoving positions ${\bf x} = (x_1, x_2, x_3)$ of halo centers
in the $(x_1, x_2)$ coordinate plane of the simulation box. 
To avoid overcrowding, we consider only halo centers that satisfy 
the condition $0\, \leq \,x_3 \,\leq \,5\hmpc$. Halos with masses exceeding
$\, {\tt M} = 5\cdot 10^{12}h^{-1}M_{\odot}\,$ are plotted as circles with diameters
proportional to halo masses. For reference, the size of a circle, representing a halo with a mass of
$2\cdot 10^{13}M_{\odot}$ is shown in Figure \ref{fig:halos25}, above the upper left 
frame. The halos with masses $M < {\tt M}$ are plotted as dots. It is interesting to note that
as expected, the scalar interactions enhance the ability of the larger halos to accrete
matter and become even larger. This effect is particularly striking when we compare the
upper left, Newtonian frame, with the lower right frame where
$\, r_s = 1\hmpc$ and $\, \beta = 1$. The halos seen in the rectangle
\beqa
5\hmpc \; < \; x_1 \; < 10\hmpc \;\; ,\\
15\hmpc \; < \; x_2 \; < 25\hmpc \;\; ,
\label{eqn:rectangle}
\eeqa
at lower right have already
acreted all debris in their vicinity at earlier times, while in the Newtonian frame the
accretion process is still going on. At the same time, the small-scale mass redistribution process
does not affect the large scale clustering: the cosmic web is clearly
visible in all of the frames.

\subsubsection{Low-mass Halos}

\begin{figure*}
\resizebox{120mm}{!}{\includegraphics[angle=-90]{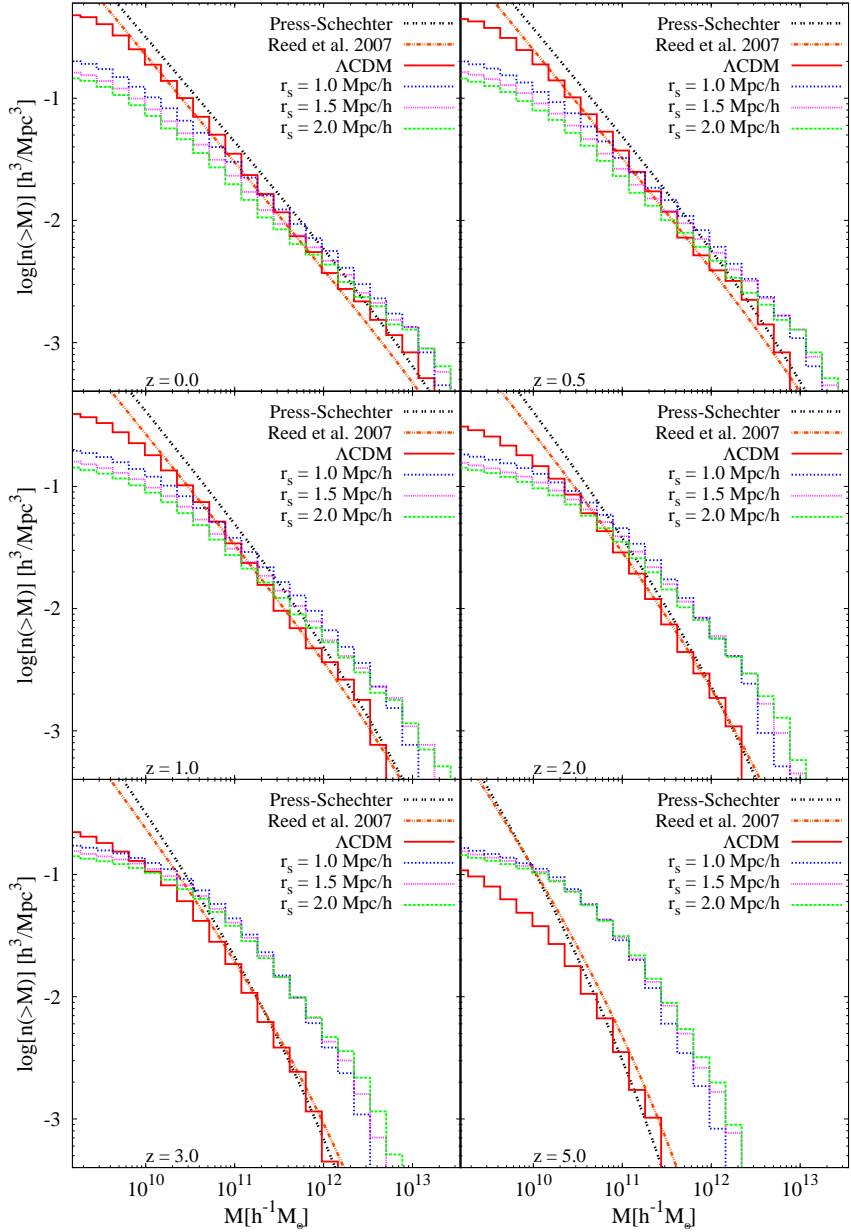}}
\caption{The redshift evolution of the cumulative halo mass functions in simulations with $\, L = 25\hmpc\,$.
We show the $\lcdm$ case and two scalar models with $\, \beta = 1\,$ and two different values of $\, r_s\,$.  
\label{fig:cmf25}}
\end{figure*}
In this section we look at the low-mass end of the halo population. 
Our Figures \ref{fig:lm_halos25} and \ref{fig:halos25} are complementary to each other.
Both are derived from the same simulation, except this time
we show only halos with masses $\, M \leq {\tt m} = 2\cdot 10^{10}h^{-1}M_{\odot}\, $.  The circles 
show halos with masses $\, 8\cdot10^9h^{-1}M_{\odot} \leq M \leq {\tt m}$. The remaining halos, with
masses $\, M \leq 8\cdot10^9h^{-1}M_{\odot}$, are plotted as dots. As before, the diameters of the
circles are proportional to halo masses. The influence of the scalar field is particularly well pronounced
when we compare frames with scalar field switched off ($\, \beta = 0$, upper left) and on
($\, \beta =1$, lower right). The low abundance of light halos, seen here, is consistent with the
high abundance of heavy halos in Figure \ref{fig:halos25}. Both figures show the
rapid accretion and massive halo formation at high redshift, induced by the scalar force.
\begin{figure*}
\resizebox{201mm}{!}
{\includegraphics[angle=-90]{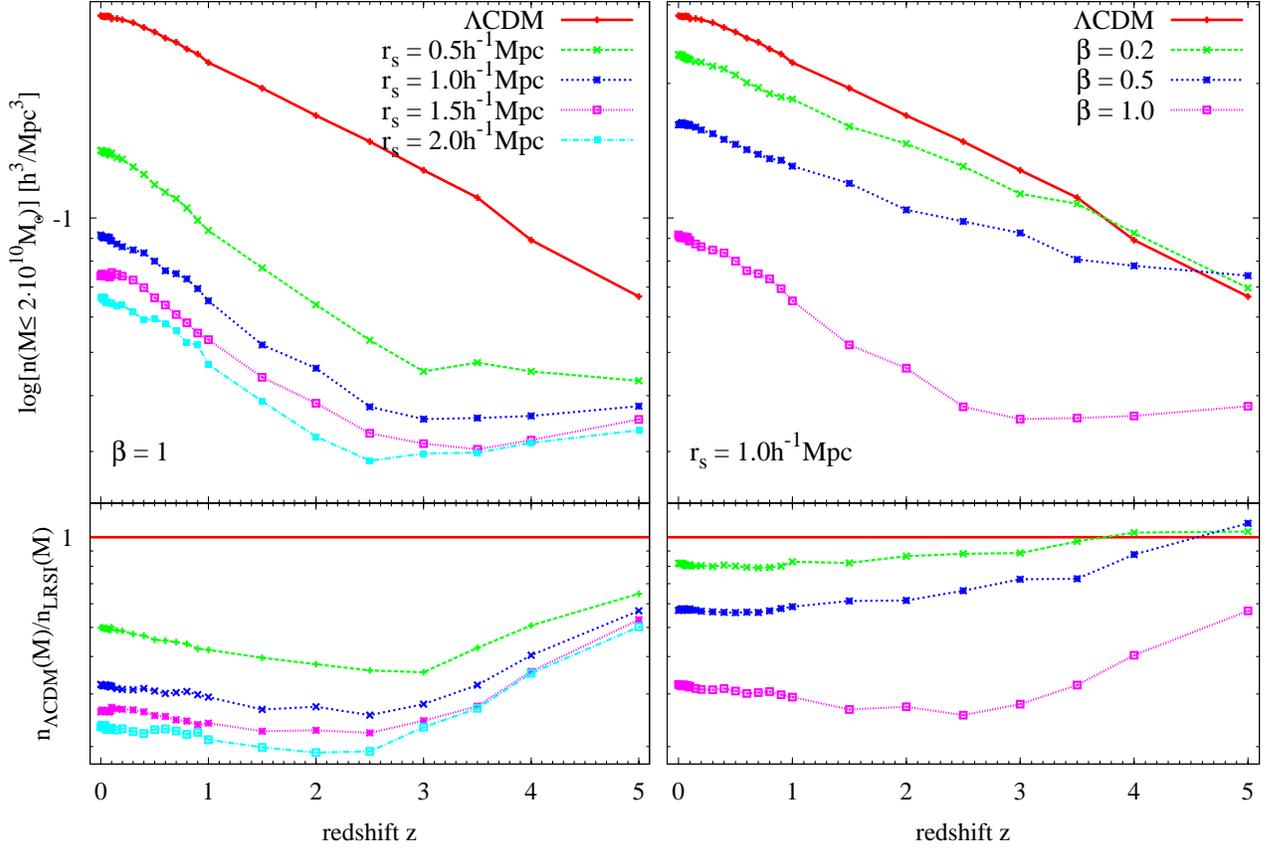}}
\caption{Redshift evolution of the abundance of low-massive halos. Here, we consider only halos with $M\leq 2\cdot 10^{10}h^{-1}M_{\odot}$ The bottom panels show the ratio of the halo abundance to the $\lcdm$ case.
\label{fig:halo_abundances1}}
\end{figure*}
\begin{figure*}
\resizebox{201mm}{!}
{\includegraphics[angle=-90]{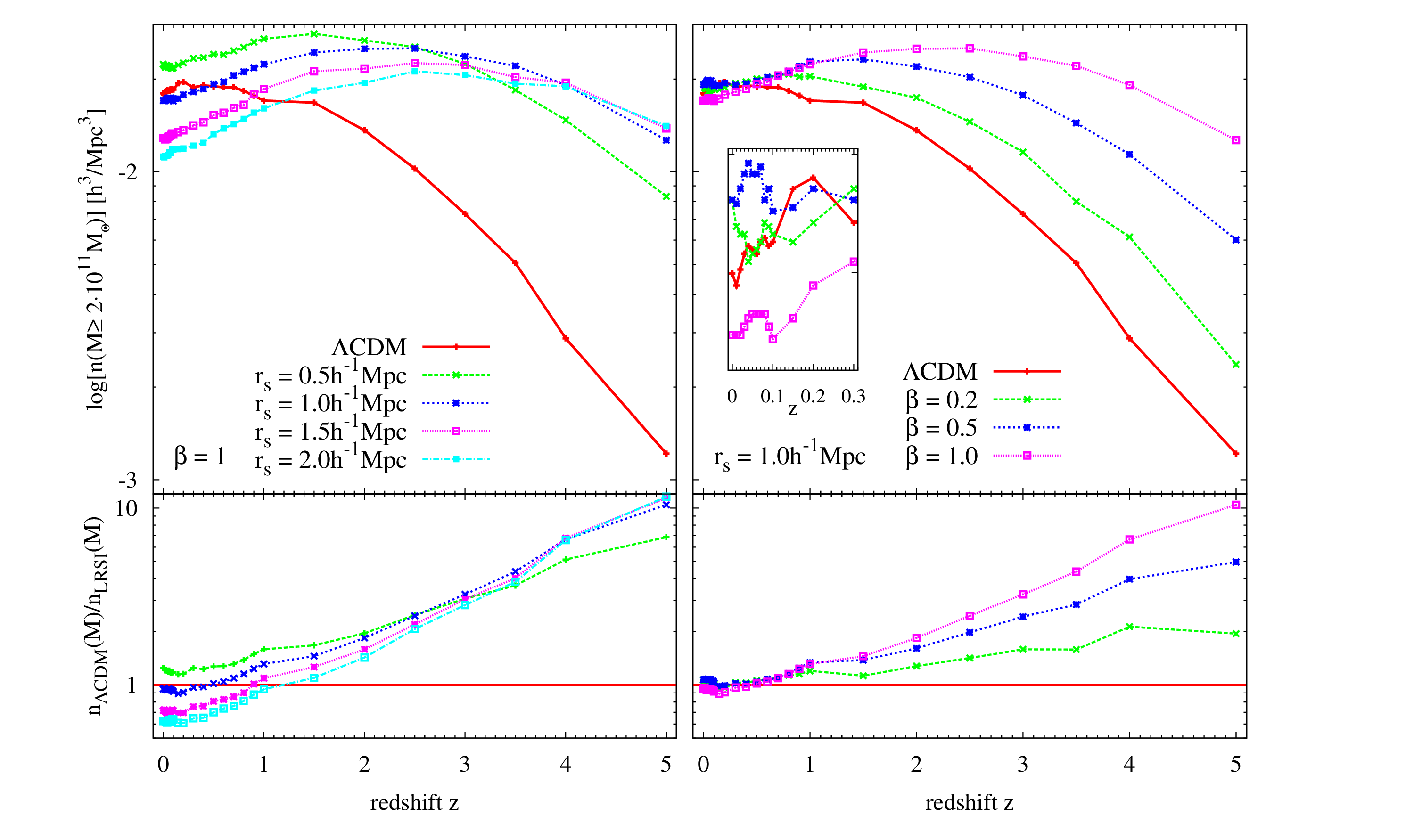}}
\caption{Redshift evolution of the abundance of high-massive halos. Here we consider only halos with $M\geq 2\cdot 10^{11}h^{-1}M_{\odot}$ The bottom panels show the ratio of the halo abundance to the $\lcdm$ case. The small box inset in the bottom right frame is a magnification
of the $\, 0\leq z \leq 0.3 \,$ region in the same frame. 
\label{fig:halo_abundances2}}
\end{figure*}

\subsubsection{The Cumulative Mass Function}

For a more quantitative description of the halo formation process, we will now introduce
the cumulative mass function 
(cmf), defined as the mean number density of halos with masses greater than the argument mass, 
\beq
n(>M) \; = \; {N_H(>M)\over L^3} \; ,
\eeq
where $N_H(>M)$ is the number of halos with masses greater than $M$, identified within
the a comoving volume $\, L^3$.  Later we will also consider the mean number density
of halos below a certain mass threshold, $\, n(<M)\,$. The sum of these two densities, 
multiplied by $L^3$ gives the mean number of halos of all masses.

In Figure \ref{fig:cmf25} we show the 
redshift evolution of the cmf for several models. We have also plotted a theoretical predictions for the $\lcdm$ cmf from the Press-Schechter formalism \cite{Press-Schechter}
and Reed \textit{et al.} \cite{Reed} using publicly available fitting formulas kindly provided by Darren Reed.
The comparison of the $\lcdm$ cmf with its scalar counterparts confirms the results
obtained by plotting the positions of halos with different masses. All scalar models show enhanced abundances
of high mass halos and reduced abundances of
low-mass halos at $\, z = 0$. For $\, r_s = 2\hmpc$, the low-mass tail of the cmf drops by almost an order of magnitude
below the Newtonian cmf. We can also notice that $\lcdm$ halo abundance seems to be underdeveloped compared to scalar-induced cmf's at high redshifts. 
This clearly implies, in our opinion, that structure formation process in LRSI models is much more efficient at high redshifts compared to the Newtonian case.
The $z=0$ frame in Figure \ref{fig:cmf25} 
shows another interesting property of the models we consider here. For halo masses in the galactic range,
$\, 10^{11}M_{\odot}\,$ to $\, 10^{12}M_{\odot}\,$, the scalar-induced halo abundances agree with the Newtonian
predictions. This is an advantage because the galactic number densities predicted by the 
$\Lambda$CDM model agree with observations \cite{2005ApJ...619..697A}. 

Another interesting phenomenon is the overproduction of high mass halos in the cluster mass range,  
$\, M \sim 10^{13}h^{-1}M_{\odot}$. The cluster abundance at $\, z =0$ is relatively well known from observations,
and it provides a strong cosmological test. It has been used in the past to exclude the once popular Einstein-de Sitter
CDM model \cite{Peebles1989}. To decide how deadly this may become for scalar interaction models, 
we need a larger box to sample the cluster population properly and more particles to keep
a reasonable force resolution. Clusters are rare objects. In a $\, 25\hmpc$ box, at the cluster mass range, we are
dominated by small-number statistics. The mean distance between a pair of
$\Lambda$CDM clusters (as well as real clusters in galaxy surveys) 
is $\sim 50 \hmpc$. Nusser \textit{et al.}, who used a $\, 50\hmpc$ box, found only a small excess of
the scalar-induced cluster halo abundance over the Newtonian case.  We plan to study this problem,
using higher resolution simulations in the near future.

Apart from plotting $\,n(>M)\,$ as a function of $M$ for fixed redshifts, it is also interesting to 
fix the mass and see how the cmf evolves with $\,z$. In Figures \ref{fig:halo_abundances1} and  \ref{fig:halo_abundances2}
we present the redshift evolution of two measures of halo abundances, 
$\,n(M\leq 2\cdot 10^{10}h^{-1}M_{\odot})$, shown in the former figure,
and $\, n(M\geq 2\cdot 10^{11}h^{-1}M_{\odot})$, shown in the latter figure. 

Note that for all scalar models, the
low-mass halo abundances are lower than in the Newtonian case at high redshift. This happens
because the enhanced gravity at small scales speeds up the halo formation process, hence also increase the typical halo mass at high redshifts compared to the $\lcdm$ case. 
The light halos become extinct earlier, because they merge with larger mass halos. 
This process is responsible for evacuating the voids more effectively than conventional Newtonian forces.

On the high mass end, we see two interesting effects:

First, because of faster accretion, the higher mass halos
reach abundances, close to their final values at relatively high redshifts. Later, their number
densities change less rapidly with redshift than the $\lcdm$ halo number density. This is particularly
prominent in the right frame at the top of Figure \ref{fig:halo_abundances2}. 
Such a picture is consistent with the observational evidence for an uneventful recent past of galaxies 
like our Milky Way and other nearby galaxies, allowing merger events only at high redshifts
\cite{M31mergers, MWquiet, 2008ApJ...681.1089R}.
In contrast, for the Newtonian model, we see rapid
growth of $\, n(M\geq 2\cdot 10^{11}h^{-1}M_{\odot})$ with decreasing redshift, suggesting that
mergers continue to the present day. 

The second effect, particularly pronounced in the right frame at the top of Figure \ref{fig:halo_abundances2},
is the convergence of all of the $\, n(>M)$ curves at $\, z = 0$. Since our mass threshhold for
this set of cumulative mass functions is in the galactic mass range, this convergence is a promising
feature of scalar models, because the $\Lambda$CDM abundance of galaxies agrees with observations \cite{Zackrisson2006}.  

\section{\label{CR}Summary}

We have performed N-body simulations of large scale structure formation in a $\Lambda$CDM background, with
a Newtonian potential and a Newtonian force law, modified by the scalar interaction. We have studied the spatial distribution of dark
matter particles and halos. We also investigated statistical measures of clustering, such as the two-point
correlation function, the power spectrum, density probability distribution function and the cumulative halo mass function. We find that the scalar
interaction removes debris from cosmic voids more effectively than the standard $\Lambda$CDM model.
It also suppresses late accretion and merger activity; halo formation processes move to higher redshifts. These
findings agree very well with earlier work \cite{NGP}. For the first time in the literature, we have
also shown the effect of scalar forces on the evolution of the power spectrum. We have resolved the
boundary between pure Newtonian dynamics, and enhanced power, generated by the scalar interactions.

In the near future we plan to run higher resolution simulations. We will follow the evolution of the baryon
spatial distribution, as well as the dark matter. We will also consider the three-point
correlation function and bispectrum - statistics of higher order, than those considered here. Finally
we expect to constrain the scalar field parameter range by direct comparisons of model predictions 
with observations.  

\acknowledgments
This research was partially supported by the Polish Ministry of Science Grant no. NN203 394234. WAH would like to thank Anatoly Klypin, Alexander Knebe, Pawel Ciecielag, Michal Chodorowski and Radek Wojtak for valuable discussions and comments. We also like to acknowledge anonymous referee who helped improve the scientific value of this article. Simulations presented in this work were performed on the 'PSK' cluster at the Nicolaus Copernicus Astronomical Center and on the 'halo' cluster at Warsaw University Interdisciplinary Center for Mathematical and Computational Modeling.

\end{document}